\documentclass[aps,floats,epsf,psfig,plb,preprint,graphics]{revtex4-1}
\usepackage{graphicx}
\usepackage{wrapfig}
\usepackage{appendix}
\usepackage{rotating}
\def\ssp{\def\baselinestretch{1.0}\large\normalsize}

\begin{document}

\input epsf.tex

\ssp

\setcounter{page}{1}

\newcommand{\mitt}{Laboratory for Nuclear Science, Massachusetts Institute of Technology, Cambridge MA}
\newcommand{\ucsb}{Department of Physics, University of California, Santa Barbara, CA}
\newcommand{\uw}{Center for Experimental Nuclear Physics and Astrophysics, University of Washington, Seattle WA}
\newcommand{\pnnl}{Pacific Northwest National Laboratory, Richland, WA}
\newcommand{\haystack}{Haystack Observatory, Massachusetts Institute of Technology, Westford, MA}
\newcommand{\caltech}{California Institute of Technology, Pasadena, CA}
\newcommand{\nrao}{National Radio Astronomy Observatory}
\newcommand{\kit}{Karlsruhe Institute of Technology, Karlsruhe, Germany}

\title{{\bf Project 8: Determining neutrino mass from tritium beta decay using a frequency-based method  \\}}

\author{P.~J.~Doe}\affiliation{\uw}
\author{J.~Kofron}\affiliation{\uw}
\author{E.~L.~McBride}\affiliation{\uw}
\author{S.~Doelman}\affiliation{\haystack}
\author{J.~A.~Formaggio}\affiliation{\mitt}
\author{D.~Furse}\affiliation{\mitt}
\author{B.~H.~LaRoque}\affiliation{\ucsb}
\author{M.~Leber}\affiliation{\ucsb}
\author{B.~Monreal}\affiliation{\ucsb}
\author{N.~S.~Oblath}\affiliation{\mitt}
\author{R.~G.~H.~Robertson}\affiliation{\uw}
\author{D.~M.~Asner}\affiliation{\pnnl}
\author{A.~M.~Jones}\affiliation{\pnnl}
\author{J.~Fernandes}\affiliation{\pnnl}
\author{A.~Rogers}\affiliation{\haystack}
\author{L.~J~Rosenberg}\affiliation{\uw}
\author{G.~Rybka}\affiliation{\uw}
\author{B.~A.~VanDevender}\affiliation{\pnnl}
\author{R.~Patterson}\affiliation{\caltech}
\author{R.~Bradley}\affiliation{\nrao}
\author{M.~Bahr}\affiliation{\ucsb}
\author{T.~Th\"{u}mmler}\affiliation{\kit}

\date{\today}

\begin{abstract}
A general description is given of Project 8, a new approach to measuring the neutrino mass scale via the beta decay of tritium.  In Project 8, the energy of electrons emitted in beta decay is determined from the frequency of cyclotron radiation emitted as the electrons spiral in a uniform magnetic field.
\end{abstract}

\maketitle

\section{Introduction}

Ever since Enrico Fermi's original proposal~\cite{fermi_tentativo_1933}, it has been known that the neutrino mass has an effect on the kinematics of beta decay.   Measurements have always suggested that this mass was very small, with successive generations of experiments~\cite{weinheimer_high_1999,Kraus:2004zw,Aseev:2011dq}  giving  a consensus upper limit $m_\beta < 2$ eV.  The upcoming KATRIN tritium beta decay experiment~\cite{Angrik:2005ep,Drexlin:2013lha,Robertson:2013aaaa} anticipates having a sensitivity of 0.20 eV at 90\% confidence.  If the neutrino mass is much below 0.20 eV, it is difficult to envision any classical spectrometer being able to access it.  Oscillation experiments, however, tell us that beta decay neutrinos are an admixture of all three mass states, at least two of which have a nonzero mass, such that the effective mass must satisfy $m_\beta > 0.01$ eV in the normal hierarchy or  $m_\beta > 0.05$ eV in the inverted hierarchy.  These bounds provide a strong motivation to find new, more sensitive ways to measure the tritium beta decay spectrum if the question of the neutrino hierarchy is ever to be resolved.

The fact that neutrinos have non-zero mass is a powerful reminder that our standard model of nuclear and particle physics remains incomplete. Direct measurements of the neutrino mass can provide direction as to how to extend that model.  Its implications are  not just limited to the field of nuclear physics, but extend equally to particle physics and cosmology. There are many theories beyond the Standard Model that explore the origins of neutrino masses and mixing. In these theories, which often work within the framework of supersymmetry, neutrinos naturally acquire small but finite masses. Several models use the so-called see-saw effect to generate neutrino masses. Other classes of theories are based on completely different possible origins of neutrino masses, such as radiative corrections arising from an extended Higgs sector. As neutrino masses are much smaller than the masses of the other fermions, the knowledge of the magnitude of neutrino masses is crucial for our understanding of the fermion masses in general. It has been pointed out  that the absolute mass scale of neutrinos may be even more significant and straightforward for the fundamental theory of fermion masses than the determination of the neutrino mixing angles and CP-violating phases~\cite{farzan_effective_2002}. It will likely be the absolute mass scale of neutrinos that will determine the scale of new physics.  For a gauge of the impact that neutrino masses can have on particle physics and cosmology, see Table~\ref{tab:knowledge}.

\begin{table*}[htdp]
\caption{Impact of neutrino mass sensitivity level as obtained from beta decay measurements on nuclear physics and cosmology.}
\begin{center}
\begin{tabular}{|l|l|l|}
\hline
Neutrino Mass Sensitivity & Scale & Impact \\
\hline
$m_\nu < 2$ eV & eV & Neutrinos ruled out as primary dark matter \\
\hline
$m_\nu > 0.2$ eV &  Degeneracy &  Cosmology, $0\nu\beta\beta$ reach \\
\hline
$m_\nu > 0.05$ eV &  Inverted Hierarchy & Resolve hierarchy if null result \\
\hline
$m_\nu > 0.01$ eV &  Normal Hierarchy& Oscillation limit, possible relic neutrino sensitivity\\
\hline
\end{tabular}
\end{center}
\label{tab:knowledge}
\end{table*}%

\section{Description of the Technique}

The most sensitive direct searches for the electron neutrino mass
up to now are based on the investigation of the electron spectrum
of tritium $\beta$-decay. As both the matrix elements and Coulomb correction are independent
of $m_\nu$, the dependence of the spectral shape on $m_\nu$ is
given by the phase space factor only. In addition, 
 neutrino mass determined from tritium $\beta$-decay is independent of whether
the electron neutrino is a Majorana or a Dirac particle.

To make advances toward lower and lower masses, it is important to develop techniques that allow for extremely precise spectroscopy of low energy electrons.  Current electromagnetic techniques, such as those employed by the KATRIN experiment, can achieve of order $10^{-5}$ in precision, but are at the limit of their sensitivity.  Should the mass prove to be beyond the reach of KATRIN, a new technique must be pursued in order to approach the inverted or even the hierarchical neutrino mass scale implied by current oscillation measurements.  The technique proposed here relies on the principle that the frequency of cyclotron radiation emitted by a relativistic electron depends inversely on its energy, independent of the electron's direction when emitted.  As the technique inherently involves the measurement of a {\em frequency} in a non-destructive manner, it can, in principle, achieve a high degree of resolution and accuracy.  The combination of these two features makes the technique attractive within the context of neutrino mass measurements.

A charged particle, such as an electron created from the decay of tritium or from neutrino capture, traveling in a uniform magnetic field $B$ in the absence of any electric fields,  will travel along the magnetic field lines undergoing simple cyclotron motion.  The characteristic frequency $\omega$ at which it orbits is given by
\begin{equation}
\label{eq:cyfreq}
\omega = \frac{e B}{\gamma m_e} = \frac{\omega_c}{\gamma} = \frac{\omega_c}{1+\frac{E}{m_e c^2}},
\end{equation}
\noindent where $\omega_c$ is the cyclotron frequency, $E$ and $m_e$ are the electron kinetic energy and mass, respectively, and $\gamma$ is the relativistic boost factor. The cyclotron frequency, therefore, is shifted according to the kinetic energy of the particle and, consequently, any measurement of this frequency stands as a measurement of the electron energy.  Electrons from tritium decay have a kinetic energy of 18.6 keV or, equivalently, a boost factor $\gamma \simeq 1.0364$.

A charged particle undergoing cyclotron motion will also emit cyclotron radiation as it travels through a magnetic field.  Since the relativistic boost for the energies being considered is close to unity, the radiation emitted is relatively coherent.  For a magnetic field strength of 1 Tesla, the emitted radiation has a frequency of 27 GHz.  This frequency band is well within the range of  commercially available radio-frequency antennas and detectors. It is conceivable, therefore, to make use of radio-frequency (RF) detection techniques in order to achieve precision spectroscopy of electrons.  Furthermore, the typical power emitted by these electrons is sufficiently high to enable single-electron detection. A more in-depth description of the technique, including a discussion of its potential sensitivity, can be found in Ref~\cite{Monreal:2009za}.

\section{Sensitivity to Neutrino Mass}

For an electron-flavor-weighted neutrino mass $m_\nu = m_\beta$ that is not too small, the tritium beta spectrum can be written in a simplified form,
\begin{eqnarray}
\frac{dN}{dE_e} &=& 3rt(E_0-E)[(E_0-E)^2-m_\nu^2]^{1/2}
\end{eqnarray}
where $r$ is the rate in the last eV of the spectrum in the absence of mass, t is the running time and $E_0$ is the endpoint energy.   The neutrino mass can be determined from a single measurement of the number of events in an interval $\Delta E = E_0-E_1$ from the endpoint energy, as long as other parameters, namely the rate, time, endpoint energy, and background, are well enough known.  This is an idealization but  not unrealistic for an experiment like Project 8 where very high statistics data on background and the spectrum below the endpoint are automatically taken ``for free'' because all events are recorded as they occur.  (KATRIN takes data point--by-point.)  One can do still better statistically by gaining more information about the distribution of events within the window $\Delta E$, but this ansatz provides a robust statistical baseline for estimating the precision that can be obtained.  It is not assumed that the endpoint energy is known in an absolute sense; it need only be determined relative to the analysis window.

The total number of signal events $N_s$ in this window is obtained by integrating, 
\begin{eqnarray}
N_s &=& rt (\Delta E)^3\left[1-\frac{3}{2}\frac{m_\nu^2}{(\Delta E)^2}\right].
\end{eqnarray}
If there is in addition a background $b$ that is energy-independent and proportional to the width $\Delta E$ of analysis window, the total number of events is 
\begin{eqnarray}
N_{\rm tot} &=& rt (\Delta E)^3\left[1-\frac{3}{2}\frac{m_\nu^2}{(\Delta E)^2}\right] +bt\Delta E
\end{eqnarray}
The statistical uncertainty $\sigma_{m_\nu^2} $ is thus related to the variance in the total number of events:
\begin{eqnarray}
\sigma_N^2 &=& \left(\frac{\partial N_{\rm tot}}{\partial m_\nu^2}\right)^2 \sigma^2_{m_\nu^2} \\
\frac{\partial N_{\rm tot}}{\partial m_\nu^2} &=& -\frac{3rt\Delta E}{2} \\
\sigma_{m_\nu^2} &=& \frac{2}{3rt\Delta E}\sqrt{N_{\rm tot}} \\
 &= & \frac{2}{3rt}\sqrt{rt\Delta E + \frac{bt}{\Delta E}},  \label{eq:sig}
\end{eqnarray}
There is an optimum choice of $\Delta E$ that minimizes the uncertainty,
\begin{eqnarray}
\Delta E_{\rm opt} &=& \sqrt{\frac{b}{r}}. 
\end{eqnarray}
The minimum is broad.  Setting the analysis window incorrectly by a factor $m$ results in an increase in the statistical uncertainty by a factor
\begin{eqnarray}
\frac{\sigma}{\sigma_{\rm opt}} &=& \sqrt{\frac{1}{2}\left(m+\frac{1}{m}\right)}.
\end{eqnarray}
If $\Delta E$ is off by a factor of 2 from the optimum, there is a 10\% increase in the statistical uncertainty.  As a practical matter, it may not always be possible to achieve an instrumental width of the optimum size when rates are high or backgrounds low.  Moreover, improving the instrumental resolution beyond a certain point is not useful because there is a limit set by the broadening caused by the final-state spectrum (FSS) in the decay of molecular T$_2$ to T$^3$He$^+$.   The FWHM of this distribution is about 0.7 eV \cite{PhysRevLett.84.242}.  The instrumental resolution itself has two readily identifiable components, the field inhomogeneity and the finite duration of a cyclotron-emission wavetrain before the electron scatters.  To allow for these contributions, we adopt a composite analysis window width 
\begin{eqnarray}
\Delta E &=& \sqrt{ \frac{b}{r} +( \Delta E_{\rm FSS})^2+\left(\frac{E}{(\gamma-1)} \frac{2.35 \sigma(B)}{B}\right)^2+\left(\frac{E}{(\gamma-1)} \frac{2.35 \beta c \sigma_0 n}{2\pi f_c}\right)^2 } \label{eqten}
\end{eqnarray}
where $\Delta E_{\rm FSS}$ is a minimum useful width set by final-state broadening, $\sigma(B)$ is the rms field variation in the active region, $\sigma_0$ is the scattering cross section per molecule, $n$ is the number of  molecules per unit volume, and $f_c$ is the cyclotron frequency.  The contribution due to scattering is neither Gaussian nor Lorentzian when short wavetrains are rejected as they would be experimentally; a Gaussian is assumed here for convenience.   

The decay rate R in a volume V is related to the number density $n$ through the mean lifetime $\tau_m$,  
\begin{eqnarray}
R &=& n \frac{ V}{\tau_m}.
\end{eqnarray}
Electrons with a shallow pitch angle are not trapped, introducing a solid angle $\Delta\Omega$.  Other contributions to inefficiency can be merged and the solid angle becomes a general efficiency.  Hence the detected decay rate 
\begin{eqnarray}
r &=& \Delta\Omega \frac{n V}{\tau_m } \eta,
\end{eqnarray}
where $\eta$ is the branching ratio to the uppermost 1 eV of the spectrum ($\sim 2 \times 10^{-13}$).

The fact that the last  contribution to the instrumental width in Eq.~\ref{eqten} is one that scales with the total rate introduces a surprising effect.  There is no longer an optimum choice for $\Delta E$.  Instead, the statistical precision improves slowly but steadily for increasing values of the density $n$.  But it is not an applicable strategy because a collision-dominated resolution function cannot be much broader than the neutrino mass effect sought.  Systematic uncertainty in the resolution will limit the density.

There is a simple relationship between the uncertainty in the variance of an instrumental resolution contribution and the corresponding uncertainty introduced in the neutrino mass:
\begin{eqnarray}
\sigma_{m_\nu^2} &\approx & 2  \sigma_{\rm res}^2.
\end{eqnarray}
Each of the resolution components in Eq.~\ref{eqten} has an associated uncertainty that propagates into the neutrino mass.   For concreteness, we assume that the distributions are each  known to 1\%.

\begin{figure}[htb]
   \begin{center}
   \includegraphics[width=6in]{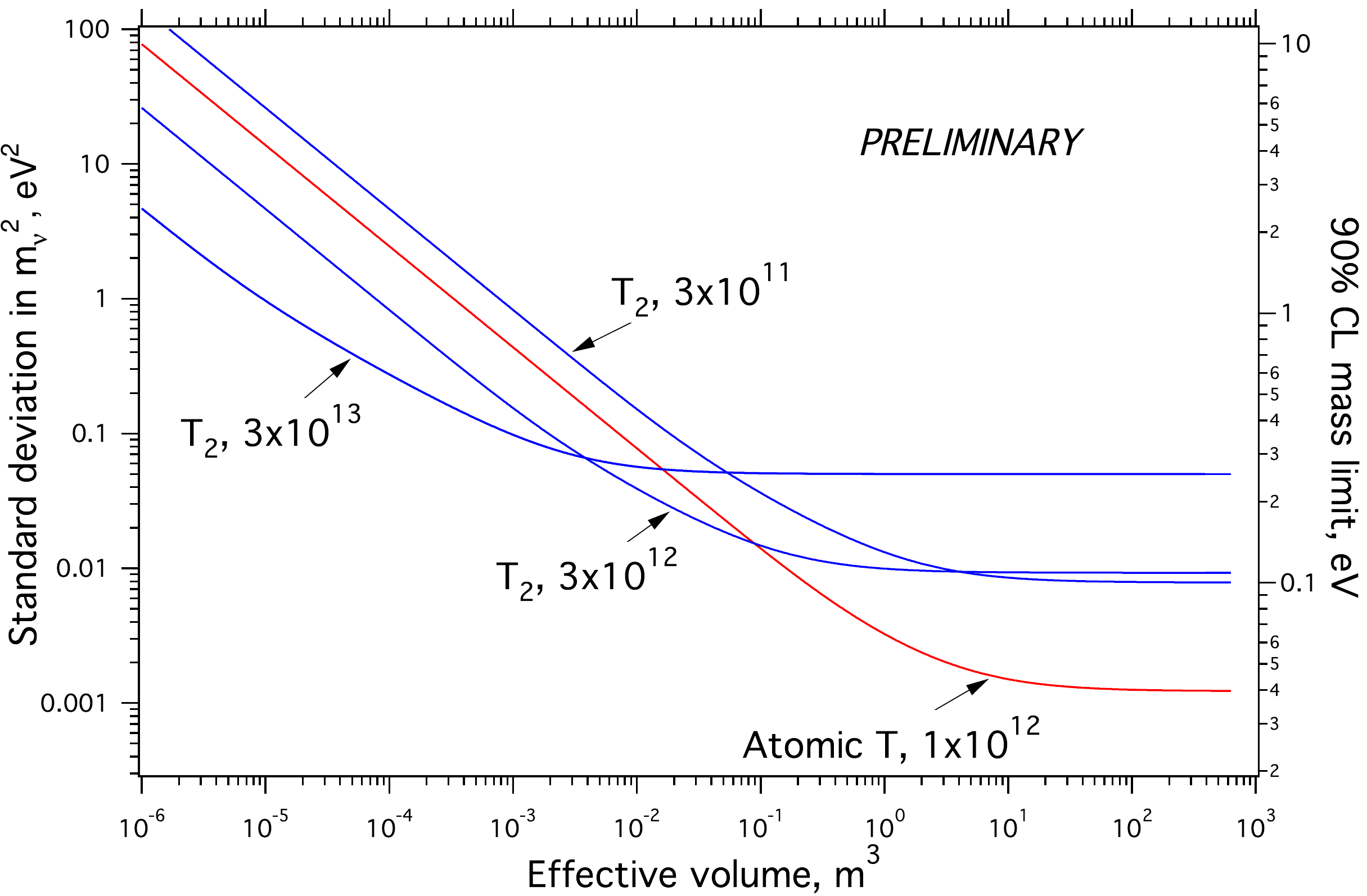}
   \caption[]{Uncertainty obtainable as a function of volume under observation for various choices of number density per cm$^3$.  Systematic uncertainties due to imperfect knowledge of contributions to the resolution are included.  The frequency chosen is 26.5 GHz, the field is uniform to 0.1 ppm rms, the source temperature for molecular T$_2$ is 30K and for atomic T it is 1K, and the background is $10^{-6}$ per second per eV.  The efficiency factor $ \Delta\Omega$ is taken as unity for the effective volume, and the live time is $3\times10^7$ seconds.}
   \end{center}
   \label{fig:sensitivity}
\end{figure}
Figure 1   shows calculated neutrino mass statistical and systematic sensitivities for various choices of number density, as a function of  volume.  The cross-section cited by Aseev {\em et al.}~\cite{aseev_energy_2000}, $3.4\times 10^{-18}$ cm$^2$, has been used for electron scattering by molecules, and for atoms we have used $9\times 10^{-19}$ cm$^2$ based on the work of Shah {\em et al.}~\cite{shah87}.  For calculating the `sensitivity' shown here, the expected value for ${m_\nu^2}$ is taken to be 0, and, statistically, positive and negative values for this quantity are equally probable.  The 90\% CL is a one-sided interval derived by setting the 1.28-sigma upper threshold on ${m_\nu^2}$, which is assumed to be Gaussian distributed.  The square root of this number is displayed on the right-hand axis.  

As can be seen, an experiment with gaseous molecular T$_2$ reaches a limit in sensitivity of order 100 meV because of the width of the FSS combined with Doppler broadening associated with the minimum feasible operating temperature near 30 K.  For this reason, the Project 8 collaboration is developing an atomic T source in a magnetic configuration that traps both spin-polarized atoms and the betas.  The density required is in an achievable range, and the operating temperature needed is near 1 K.   

The physics reach of a Project 8 experiment depicted in Fig. 1 is attractive, but should be regarded as about the best that could be done with this type of measurement.  The systematic uncertainties assumed on resolution-like parameters are small and a number of presumably less important effects are omitted. 

\section{Project 8: A Multiphase Approach}

The technique described above represents a potentially novel and effective approach for measuring the energy of electrons. Some of the advantages are enumerated below: 

\begin{enumerate}
\item{\it Source = Detector:}  Since the energy measurement of the electron is non-destructive, it takes place anywhere along the path of the electron.  This feature, in combination with the transparency of the gas to microwave photons, removes the necessity of extracting the electron  from the  source in order to measure its energy .  The combination of the source and detector region as one allows for a more favorable scaling of the experiment.
\item{\it Frequency Measurement:} Frequency techniques number among the most precise and accurate types of measurement that can be made.  The linearity offered by frequency techniques allows for exquisite calibration of these measurements.  The level of precision envisioned for our measurements (of order part per million) can be achieved with standard, commercially available technology.
\item{\it Full Spectrum Sampling:} Unlike previous techniques used in beta decay experiments, the beta decay spectrum is available within a single measurement.  No scanning or integrating of the spectrum is necessary for the measurement.  This counting provides a large increase in the statistical efficiency of the experiment.
\end{enumerate}

For all the advantages offered by the above technique, the Project 8 collaboration realizes that there also exist a number of significant challenges in order to transform the concept into a competitive measurement of the neutrino mass.  The Project 8 collaboration is thus moving forward with a multiple-phased approach;  each stage providing both the necessary R\&D and key physics measurements of interest to the physics community.  Phase I establishes a proof-of-principle measurement for the cyclotron emission of energetic electrons by using $^{83m}$Kr as its electron source.  The prototype currently being assembled at the University of Washington incorporates  the main features of the envisioned full-scale experiment: a gaseous electron source, a magnetic trapping region, and the RF detection and amplification scheme \cite{Monreal:2012zz,Oblath:2011ne}.  

Future phases will shift the physics goals from proof-of-principle to  neutrino mass measurements of increasing capability, with a final goal of 50 meV in sensitivity using an atomic source. Reaching this ultimate sensitivity could address the question of neutrino hierarchy: if the observable (i.e., electron flavor) mass is less than this limit, then the hierarchy is normal and hence resolved.

\bibliographystyle{apsrev}
\bibliography{testbib}

\end{document}